\documentclass[aps,prl,twocolumn,groupedaddress,showpacs]{revtex4}

\usepackage{graphicx}
\usepackage{amsmath}
\usepackage{amssymb}
\usepackage{bm}

\begin{document}
\title{Probing number squeezing of ultracold atoms across the superfluid-Mott
insulator transition}
\author{Fabrice Gerbier}\email[email: ]{fabrice.gerbier@lkb.ens.fr}
\author{Simon F{\"o}lling} \author{Artur Widera} \author{Olaf Mandel} \author{Immanuel
Bloch}
\affiliation{Institut f{\"u}r Physik, Johannes
Gutenberg-Universit{\"a}t, 55099 Mainz, Germany.}

\date{\today}
\begin{abstract}
The evolution of on-site number fluctuations of ultracold atoms in
optical lattices is experimentally investigated by monitoring the
suppression of spin-changing collisions across the superfluid-Mott
insulator transition. For low atom numbers, corresponding to an
average filling factor close to unity, large on-site number
fluctuations are necessary for spin-changing collisions to occur.
The continuous suppression of spin-changing collisions is thus a
direct evidence for the emergence of number-squeezed states. In
the Mott insulator regime, we find that spin-changing collisions
are suppressed until a threshold atom number, consistent with the
number where a Mott plateau with doubly-occupied sites is expected
to form.
\end{abstract}
\pacs{03.75.-b,03.75.Hh} \maketitle
%
%
One of the most fundamental signatures of the Mott insulator (MI)
transition undergone by ultracold atomic gases in optical lattices
\cite{orzel2001a,greiner2002a,stoeferle2004a,greiner2002b,jaksch1998a,javanainen1999a,kashurnikov2002a,burnett2002a,shotter2002a,roth2003a,roberts2003a,garciaripoll2004a,plimak2004a}
is a drastic change in atom number statistics. In a very shallow
lattice, ultracold bosons tend to form a Bose-Einstein condensate.
In this case, a measurement of the probability for finding $n$
atoms at a given lattice site would reveal a characteristic
Poisson distribution with large on-site fluctuations. However, for
deeper lattices, the influence of repulsive interactions, which
disfavor such fluctuations, becomes increasingly dominant and
results in the emergence of number-squeezed states with suppressed
number fluctuations. Above a critical lattice depth, the ultracold
gas enters the MI regime, where the number fluctuations almost
vanish. In experiments so far, interaction-induced number-squeezed
states were detected through the observation of increased phase
fluctuations - the canonically conjugate variable to number
fluctuations \cite{orzel2001a,greiner2002a,stoeferle2004a}, or
through an increased timescale for phase diffusion
\cite{greiner2002b}.

\begin{figure}
\includegraphics[width=8.6cm]{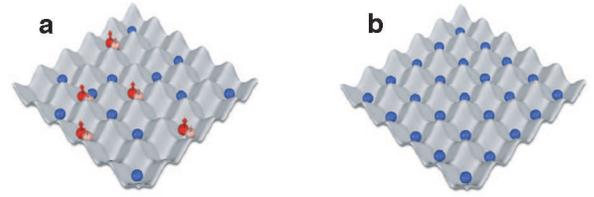}
\caption{Illustration of the number statistics measurement. Spin
changing collisions turn atom pairs initially in the Zeeman
substate $m=0$ (no arrow) to pairs in $m=\pm 1$ states (up and
down arrows). This process happens for sites with $n=2$ {\bf (a)}
or $n=3$ {\bf (b)} atoms. For one atom per site on average,
whether this occurs depends drastically on the many-body
correlations. For a Bose-Einstein condensate {\bf (c)}, large
on-site fluctuations create a finite number of sites with $2$ or
$3$ atoms, where $\pm 1$ pairs can be created. On the contrary,
for a MI state {\bf (d)}, only isolated atoms are found and no
$m=\pm1$ pairs are created.} \label{Fig1}
\end{figure}

In this Letter, we directly observe the continuous suppression of
number fluctuations when the ultracold sample evolves from the
superfluid (SF) regime to deep in the MI regime. The idea behind
our measurement is illustrated in Fig.~\ref{Fig1}. After producing
an ultracold gas in an optical lattice, we suddenly increase the
lattice intensity, suppressing tunneling and freezing the number
distribution. A probe sensitive only to the presence of {\it atom
pairs} at a given lattice site is finally applied. Close to unity
filling, a non-zero probe signal is obtained only if initially
large on-site fluctuations produce a non-zero fraction of sites
with two atoms. While we observe this behavior for a gas initially
in the SF regime, the probe signal is progressively suppressed
when approaching the Mott transition, indicating increasingly
number-squeezed states.

The specific two-particle probe used in this work are
spin-changing collisions (see \cite{widera2005a} and references
therein), which convert at each lattice site pairs of spin $f=1$
atoms in the $m=0$ Zeeman sublevel to pairs with one atom in
$m=+1$ and the other in $m=-1$. In principle, other schemes, e.g.
measuring the interaction energy \cite{roberts2003a}, or
monitoring atom losses due to Raman photoassociation
\cite{rom2004a,ryu2005a} or Feshbach resonances
\cite{stoeferle2005a}, could be suitable for this measurement.
Spin changing collisions are appealing because non destructive
(see also \cite{widera2004a}), and because they can be resonantly
controlled using the differential shift between Zeeman sublevels
induced by an off-resonant microwave field
\cite{pu2000a,sorensen2001a,gerbier2006a}. We show that this
technique allows to measure selectively doubly-occupied sites in
the optical lattice.

\begin{figure}[ht!]
\includegraphics[width=8.6cm]{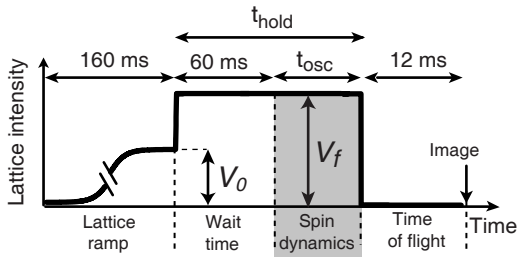}
\caption{Time sequence of the experiment.} \label{Fig2}
\end{figure}

Our experimental setup has been described in detail in
\cite{widera2005a}. We first load a degenerate gas of $^{87}$Rb
atoms in the $|F=1,m=-1\rangle$ Zeeman sublevel into a combined
magnetic trap plus optical lattice potential at an initial lattice
depth $V_0$. The intensities are then rapidly increased from $V_0$
to $V_f=40\,E_r$ within $t_{\rm up}=1\,$ms (see Fig.~\ref{Fig2}).
Here $E_r = h^2/2 M \lambda^2$ is the single photon recoil energy,
and $\lambda = 842$\,nm the lattice laser wavelength. Immediately
after this ramp, the magnetic potential is switched off, and the
cloud is held for 60\,ms in order to let the magnetic bias field
stabilize to its final value $B \approx 1.2\,$G. The atom are then
prepared in the $m=0$ state using microwave transfer pulses, and
held for a variable time $t_{\rm osc}$, during which a collisional
spin oscillation takes place. This coherent evolution is detected
experimentally as a reversible exchange between the populations in
the $m=0$ and $m=\pm 1$ Zeeman sublevels, measured by absorption
imaging after 12 ms of free expansion.

Spin changing interactions in deep optical lattices have been
described in details in \cite{widera2005a}. We consider spin $f=1$
atoms and assume that tunneling can be neglected, so that the
lattice sites are isolated from each other. At a single lattice
site, the spin-changing collisions are critically sensitive on the
filling $n$ of the well. For sites with filling $n=0$ or $1$,
spin-changing collisions cannot occur. The first non-trivial cases
corresponds to doubly- occupied wells. In this case, only two spin
states (one with both atoms in $m=0$ and the other with a single
$m=\pm 1$ pair) are accessible (see Fig.~\ref{Fig1}a). Therefore,
the atom pair undergoes Rabi-like oscillations at the effective
Rabi frequency $\sqrt{\delta_2^2+\Omega_2^2}$. The energy mismatch
(``detuning'') between the two states is
$\hbar\delta_2=\Delta\epsilon+U_s$, where
$\Delta\epsilon=\epsilon_{\rm +1}+\epsilon_{\rm -1}-2\epsilon_0$
corresponds to the difference in Zeeman energies $\epsilon_{m}$.
The spin-dependent interaction energy $U_s$ depends on atomic and
lattice parameters \cite{widera2005a}, and also determines the
coupling strength as $\Omega_2=2\sqrt{2}U_s$. Sites with $n=3$
atoms behave in a similar way (Fig.~\ref{Fig1}a), however with an
energy difference $\hbar\delta_3=\Delta\epsilon-U_s$ and a
coupling strength $\Omega_3=2\sqrt{6}U_s$.

In principle, site occupancies $n>3$, whose spin dynamics involve
more than one $m=\pm 1$ pair, are also possible. However, during
the hold time $t_{\rm hold}$ indicated in Fig.~\ref{Fig2}, those
sites can be emptied by three-body recombination (3BR) events at
an event rate $\gamma_{n}=\gamma_{\rm 3B}n(n-1)(n-2)$
\cite{jack2003a}, with $\gamma_{\rm 3B}\approx 0.5\,$s$^{-1}$ for
our parameters. Therefore, sites with $n \geq 4$ are efficiently
removed after the wait time.

In our experiment, we produce large ensembles of atoms in the
optical lattice with spatially inhomogeneous atom number
distribution. The inhomogeneity results from an additional
trapping potential $V_{\rm ext}$ present on top of the optical
lattice \cite{Vext}. In the MI regime, this potential leads to the
formation of flat Mott plateaux with well-defined atom number per
site \cite{jaksch1998a,kashurnikov2002a,batrouni2002a}. Also, the
local fluctuations have an inhomogeneous distribution.
Experimentally, we measure the ``spin-oscillation amplitude'' for
the entire atomic cloud, i.e. the global population
$\mathcal{A}_{\rm osc}=(N_{+1}+N_{-1})/N$ of the $m=\pm1$ states
after an evolution time $t_{\rm osc}$, normalized to the total
atom number $N$. This amplitude is related to the probability
$\overline{\mathcal{P}}_n$ of finding $n$ atoms per lattice site,
averaged over the cloud spatial profile.

Let us suppose that we are able to tune the single-particle
detuning to $\Delta\epsilon=-U_s$, such that doubly-occupied sites
are exactly on resonance. Then, neglecting sites with $n\geq4$,
the oscillation amplitude is obtained by summing the contribution
from sites with $n=2$ and $n=3$,
\begin{eqnarray}\label{Npm1}
\mathcal{A}_{\rm osc} \approx
\overline{\mathcal{P}}_2\sin^2\left(\frac{\Omega_2 t_{\rm
osc}}{2}\right) +\frac{6}{7}~\overline{\mathcal{P}}_3~\sin^2\left(
\sqrt{\frac{7}{8}}\Omega_2 t_{\rm osc}\right).
\end{eqnarray}
From Eq.~(\ref{Npm1}), we conclude that atom pairs and triplets
oscillate essentially out of phase. By choosing
$\Delta\epsilon=-U_s$ and $t_\pi=\pi/\Omega_2$, all
doubly-occupied sites are converted to $m=\pm 1$ pairs, whereas
the conversion efficiency for triplets is around 3 \%. Recording
the amplitude of the spin oscillations thus allow to probe the
distribution of atom pairs alone. This is reminiscent of cavity
quantum electrodynamics \cite{brune1996a,varcoe2000a}, where Fock
states of the cavity field could be discriminated due to different
coupling strengths to an atomic transition. In particular,
choosing $\Delta\epsilon=U_s$ would allow to measure the fraction
of triply-occupied sites remaining after three-body decay.

\begin{figure}
\includegraphics[width=8.6cm]{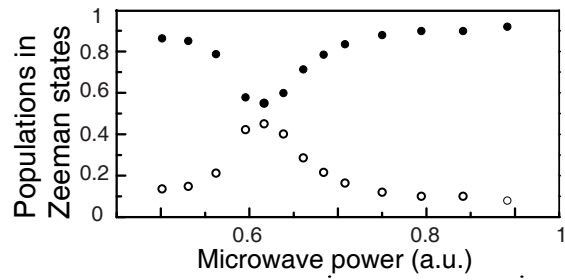}
\caption{Resonance curve of the spin-oscillation amplitude in the
far-detuned microwave field, measured at a magnetic field
$B\approx1.2~$G and at a fixed hold time $t_{\rm hold}=15.5$\,ms.
Filled (hollow) circles denote the population in the Zeeman
substate $m=0$ (resp. in $m=\pm 1$).} \label{Fig3}
\end{figure}

\begin{figure*}[ht!!!]
\includegraphics[width=18cm]{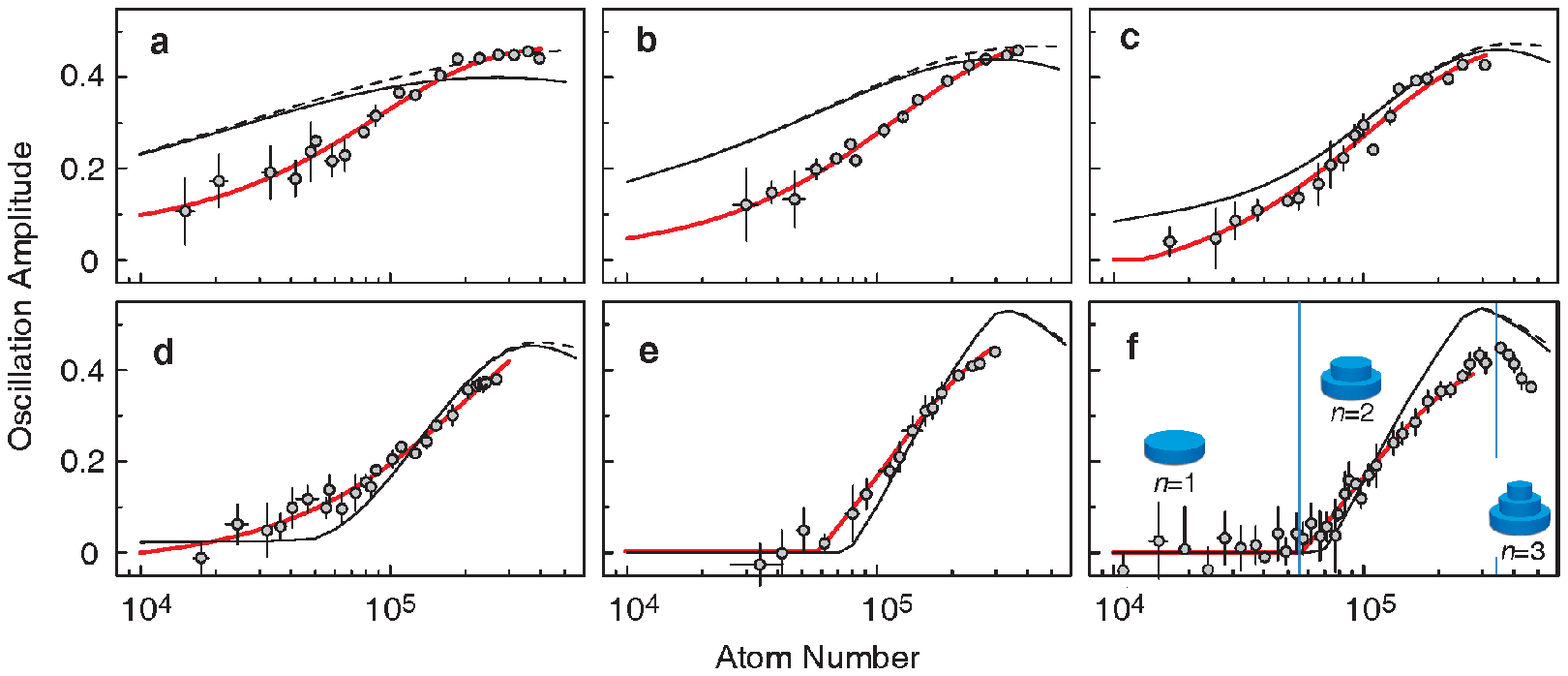}
\caption{Amplitude of the spin oscillation vs atom number and
different lattice depths: $V_0=4,8,11,13,20,40~E_r$ {\bf (a-f)}.
The thin dashed lines show the prediction of a theoretical model,
where the atom number distribution is deduced from a mean field
approach at $T=0$ (see text). The thin solid lines show only the
fraction of atom pairs calculated from the same model. The thick
solid lines are guides to the eye. In {\bf (f)}, the vertical
lines indicate where Mott plateaus with $2$ and $3$ atoms per site
are expected to form.} \label{Fig4}
\end{figure*}

To achieve full conversion of doubly-occupied sites, it is
necessary to tune the spin oscillations for doubly occupied sites
into resonance, i.e. set $\Delta \epsilon=-U_s$. In a magnetic
field $B$, the quadratic Zeeman shift contributes a {\it positive}
amount to $\Delta \epsilon$. Hence, if $U_s>0$ (which is the case
for $^{87}$Rb), the interaction energy $U_s$ leads to a residual
detuning in zero magnetic field that prevents to reach the
resonance. For this reason, we introduce a different technique
using the differential level shift induced on the individual
Zeeman sublevels by a far off-resonant microwave field
(``AC-Zeeman shift''). With a suitable choice of polarization,
detuning and power, the detuning $\Delta\epsilon$ can be tuned at
will in the range of interest, and allows to compensate the
magnetic field contribution to $\Delta\epsilon$ {\it plus} the
interaction term $U_s$. In this work, the microwave field is
detuned by several hundred MHz to the red of any hyperfine
resonance to suppress population transfer to $f=2$. Indeed, no
such transfer is observed within our experimental sensitivity.

In Fig.~\ref{Fig3}, the fraction of atoms found in $m=\pm 1$ is
plotted as a function of the microwave power for a fixed $t_{\rm
osc}=15.5\,$ms $\approx t_\pi$, corresponding to maximum
conversion. These data were taken for constant initial lattice
depth ($V_0=V_f=40\,E_r$) and atom number ($N\approx 2.6 \times
10^5$). For very low microwave powers, the spin dynamics is
suppressed by the quadratic Zeeman detuning
($\Delta\epsilon\approx2\pi\times 207~$Hz), much larger than the
spin-dependent interaction $U_s\approx 2\pi\times10.7\,$Hz. The
AC-Zeeman shift can compensate for this detuning and for the
interaction part, inducing a resonance in the number of $m=\pm 1$
pairs shown in Fig.~\ref{Fig3}. The oscillation amplitude, close
to the expected $\mathcal{A}_{\rm osc}\approx0.5$ indicates that
nearly all atom pairs are converted into $\pm 1$ pairs, in
agreement with further experiments discussed in a companion paper
\cite{gerbier2006a}.

We now turn to the measurement of number statistics. We choose
$t_{\rm osc} \approx 15.5\,$ms and and a dressing field tuned to
resonance, as in the previous paragraph. At a given lattice depth
$V_0$, we have recorded the oscillation amplitude in a broad range
of atom numbers, from about $10^4$ to a few $10^5$ \cite{Tc}. The
experiment is then repeated for various lattice depths, from the
SF regime ($V_0=4\,E_r$) to deep in the MI regime ($V_0=20\,E_r$
and $40\,E_r$). As shown in Fig.~\ref{Fig4}, at low lattice
depths, the spin oscillations occur for any atom number $N$, with
an amplitude slowly increasing with $N$. For small atom number,
the oscillation amplitude is increasingly suppressed with
increasing lattice depth, and completely vanishes for large
lattice depths. This qualitative behavior is consistent with the
behavior expected from the Bose-Hubbard model
\cite{jaksch1998a,javanainen1999a,kashurnikov2002a,burnett2002a,shotter2002a,roth2003a,roberts2003a,garciaripoll2004a,plimak2004a}.
On approaching the Mott transition, the ground state adapts to an
increased interaction energy  by reducing its number fluctuations,
eventually producing an array of one-atom Fock states at each site
where spin-changing collisions cannot occur.

Within the MI regime (Fig.~\ref{Fig4}d-f), we observe that the
suppression of spin oscillations persists up to some threshold
atom number ($6.0(3)\times10^4$ for the data in Fig.~\ref{Fig4}f).
This is consistent with the expected formation of Mott plateaus
with increasing atom number, as the cloud expands in the trapping
potential. A Mott plateau with $n$ atoms per site forms when the
cloud radius reaches the size $R_n$ where the potential energy
$V_{\rm ext}(R_n)$ matches the on-site interaction energy
$U(n-1)$. For a harmonic potential with trapping frequency
$\omega_{\rm ext}$, this happens at a threshold number
\cite{demarco2005a} $N_{n}\approx N_2\sum_{k=1}^{n} k^{3/2}$.
Above $N_2\approx4\pi/3(m\omega_{\rm ext}^2d^2/2U)^{-3/2}$, a core
with two atoms per site starts to grow, thus enabling the spin
oscillations. For the parameters that correspond to
Fig.~\ref{Fig4}f ($\omega_{\rm ext}=2\pi\times 80\,$Hz and
$V_0=40\,E_r$), we calculate $N_{\rm th}\sim6.8\times10^4$, close
to the measured value. For even higher atom number (corresponding
to $N_3 \sim 3 \times 10^5$), a shell of triply-occupied sites
start to form, reducing the fraction of atoms in the $n=2$ shell.
This can be seen in Fig.~\ref{Fig4}f, where we indeed observe a
decrease of the spin amplitude above this number.

In order to compare our experimental results with the prediction
of the Bose-Hubbard model \cite{jaksch1998a}, we solve this model
numerically within a mean-field approximation at zero temperature
\cite{sheshadri1993a,vanoosten2001a}. Accounting for losses during
the hold time $t_{\rm hold}$ (wait time plus oscillation time), we
obtain the distribution $\overline{\mathcal{P}}_n$. For each
filling $n$ and a given $t_{\rm osc}$, we calculate the conversion
efficiency $\eta_{\pm 1}(n)$ to $m=\pm 1$ pairs, and obtain the
total spin amplitude from $\mathcal{A}_{\rm osc}=\sum_n n
\eta_{\pm 1}(n)\overline{\mathcal{P}}_n$. The results of this
calculation, indicated by the solid line in Fig.~\ref{Fig4}, lie
very close to the fraction of pairs $\overline{\mathcal{P}}_2$
predicted by the same model (dashed line), in agreement with the
arguments leading to Eq.~\ref{Npm1}. Deep in the MI regime
(Fig.~\ref{Fig4}e-f), the calculations agree well with the
measurements. For lower lattice depths, although the qualitative
trend is still reproduced, we find discrepancies. Near the Mott
transition (Fig.~\ref{Fig4}d), the mean-field calculations predict
an amplitude lower than observed, a behavior consistent with the
study of number correlations beyond mean-field reported in
\cite{garciaripoll2004a}. Below the transition point
(Figs.~\ref{Fig4}a-c), the model predicts an oscillation amplitude
higher than observed. Deviations from the initial distribution may
arise in this low lattice depth regime if excitations are
generated during the preparation phase and result in an increased
populations in the low-density regions of the cloud, which barely
participate to the spin oscillations. Such ``finite temperature''
effects have possibly less influence in the MI regime, where the
many-body system is protected by an interaction gap.

In conclusion, we have shown how spin oscillations can be used to
probe number squeezing in optical lattices via the detection of
the fraction of atom pairs. Our observations confirm the expected
scenario: near-Poissonian fluctuations for shallow lattices,
strongly suppressed fluctuations for deep lattices, and a smooth
interpolation in between. Moreover, the observed behavior is
consistent with the expected formation of Mott plateaus, a
signature of the incompressibility of this system. Our results
indicate that number squeezing is robust with respect to
experimental manipulations, such as transfer to the purely optical
trap. In this sense, they are promising to employ those number
squeezed states, e.g. in Heisenberg-limited atom interferometry
\cite{dunningham2004a}.
\begin{acknowledgments}
We acknowledge support from the DFG, from AFOSR and from the EU
under the OLAQUI and the Marie Curie EIF (FG) programs.
\end{acknowledgments}
%

\begin{thebibliography}{30}
\bibitem{Vext} The potential $V_{\rm ext}$ is nearly harmonic, with a trapping frequency
$\omega_{\rm ext}\approx\sqrt{\omega_{\rm m}^2+8V_0/mw^2}$, where
$\omega_{\rm m}=2\pi\times16$\,Hz is the oscillation frequency in
the magnetic trap and where $w\approx136\,\mu$m is the laser beam
size.
\end{thebibliography}
%

%
\end{document}